# Topological Cathodes:
# Controlling the Space Charge Limit of Electron Emission Using Metamaterials

David H. Dowell[1]

SLAC National Accelerator Laboratory, 2575 Sand Hill Road, Menlo Park, California 94025, USA

**Abstract**

The space charge limit (SCL) of emission from photocathodes sets an upper limit on the performance of both high- and low-field electron guns. Generally, one is forced to strike a compromise between the space charge limit and the cathode's intrinsic emittance (I. Bazarov et al., Phys. Rev. Lett., **102**,104801(2009)). However, it is possible to nearly eliminate the SCL due to the image charge by engineering the topography of the cathode's surface. A cathode with a surface plasma frequency below the frequency spectrum of the accelerating electrons will greatly reduce the bunch's image charge or polarization of the cathode, resulting in a small image-charge field. Thereby mitigating the cathode's space charge limit. In the work presented here, a theory for the image-charge field produced by a disk of charge being accelerated from the cathode surface is developed to include the frequency-dependent behavior of surface dielectric function on the fields seen by the beam. The paper applies this theory to mitigate the SCL of a novel cathode based upon a wire-array metasurface. It is shown that such meta-cathodes should have negligible space charge limits for photoelectric, thermionic and field emission.

## I. Introduction

Our ability to assemble sub-wavelength, atom-like structures into man-made metamaterials with exotic properties is revolutionizing the fields of photonics, acoustics and the material sciences. Metamaterials consisting of 3D arrays of 'artificial atoms' in the form of miniature resonators have been shown to exhibit unnatural optical properties like negative index of refraction. Research of these new materials is beginning to make even outlandish concepts like optical cloaking and perfect lenses demonstrable possibilities.

This paper attempts to incorporate these new metamaterial ideas into beam physics and accelerator technology. Here these concepts are applied to design a cathode with the goal of mitigating the cathode's space charge limit of emission. This is done by choosing a metasurface whose plasma frequency is below the frequency spectrum of the emitted electrons' image-charge field. Such a metasurface can greatly reduce the image-charge field and thereby mitigate the space charge limit.

A smaller retarding image-charge field would lead to a significantly higher cathode field for high charge bunches. In addition, a negligible image-charge field would allow using a smaller laser spot without concern of the space charge limit (SCL), thereby reducing the beam's mean transverse energy and intrinsic emittance. A smaller beam size would also reduce the chromatic and geometric aberrations in the injector's optics.

In this work the image method is used to obtain the image-charge field near the surface of the cathode. This field is evaluated at the electron bunch location and travels with the bunch as it accelerates from the cathode. Fourier transforms are used to combine the image-charge field with the surface loss function to obtain the electric field experienced by the bunch interacting with the polarization charge on the cathode's surface. A simple wire-array metamaterial is used with the theory to explore the range of the fields generated by this bunch-cathode interaction.

[1] dowell@slac.stanford.edu; ddowell@centurylink.net

Here we ignore the space-charge forces between electrons within the bunch. These forces result from the bunch's self-energy or potential energy of assembly which drives radial and longitudinal expansion. Researchers have shown in both theory [1,2] and experiment [3] that if this expansion is linear then it produces no emittance growth.

Section II describes the connection between the image charge and the dielectric function. The relationship between the dielectric function and the surface loss function (aka. the image-to-real charge ratio) is given and the conditions for controlling the SCL are described. A thin disk model of the beam which combines the time-dependence of the image-field with the surface response function is developed in Section III. In Section IV a metasurface consisting of a rectangular array of parallel wires is used to illustrate the large range of dielectric properties a metamaterial can produce. The section also describes an optimized metasurface and discusses the practical difficulties of using structured surfaces in high electric fields. Section V contains concluding remarks with suggestions for further research.

## II. The image charge and the dielectric function of cathodes

The image method [4] gives the ratio of the image charge, q', to the real external charge, q, as

$$\frac{q'}{q} = -\left(\frac{\epsilon_c - 1}{\epsilon_c + 1}\right) \tag{1}$$

Where $\epsilon_c$ is the cathode's relative electric permittivity defined in terms of its electric permittivity, $\epsilon_{cathode}$, and electric susceptibility, $\chi_e$,

$$\epsilon_{cathode} = \epsilon_0 \epsilon_c = \epsilon_0 (1 + \chi_e) \tag{2}$$

In this paper the frequency-dependent relative permittivity, $\epsilon_c(\omega)$, is referred to as the dielectric function of the cathode.

The surface loss function at low momentum transfer [5] is,

$$g(q \to 0, \omega) = \left(\frac{\epsilon(\omega) - 1}{\epsilon(\omega) + 1}\right) \tag{3}$$

Comparing with Eqn. (1) indicates the image charge depends upon frequency and is related to the surface loss function when $q \sim 0$ for photoemission. Thus, we define the surface loss function for the cathode as

$$g_c(\omega) \equiv \left(\frac{\epsilon_c(\omega) - 1}{\epsilon_c(\omega) + 1}\right) = -\frac{q'}{q} \tag{4}$$

The negative of the image-to-real-charge charge ratio is the cathode's surface loss function.

Drude's theory [6] defines the dielectric function in terms of the material's plasma frequency and relaxation or collision time. In this theory the dielectric function is specified by two parameters, $\omega_p$ and $\tau_p$,

$$\epsilon(\omega) = 1 - \frac{\omega_p^2}{\omega\left(\omega + \frac{i}{\tau_p}\right)} \tag{5}$$

Here $\omega_p$ is the plasma frequency which is related to $n$, the number density of conduction band electrons,

$$\omega_p^2 = \frac{4\pi n e^2}{\epsilon_0 m} \tag{6}$$

and the relaxation time, $\tau_p$, depends upon both the conduction band density and the DC electrical conductivity, $\sigma$,

$$\tau_p = \frac{m\sigma}{ne^2} \tag{7}$$

where m is the mass of an electron and e is its charge. These two properties determine the magnitude and frequency behavior of q'/q and the SCL.

Table I gives the free electron properties of the commonly used cathode materials of copper, cesium and gallium. For all three elements the plasma frequency is four-orders of magnitude higher than the frequency spectrum of a 1 ps (~$10^{12}$ rad/sec) bunch. Therefore, for cathodes made from these elements, we expect to always have $\omega \ll \omega_p$ and q'/q=-1.

*Table I. Free electron properties of metals commonly used for cathodes. The electron densities are taken from Table 1.1 and the relaxation times from Table 1.3 of Ref [6].*

| | Copper | Cesium | Ga |
|---|---|---|---|
| number density, n (/m$^3$) | $8.47\times10^{28}$ | $9\times10^{27}$ (5K) | $1.54\times10^{29}$ |
| relaxation time, $\tau$ (sec) | $2.7\times10^{-14}$ | $2.1\times10^{-14}$ | $1.7\times10^{-15}$ |
| Plasma freq., $\omega_p$ (rad/s) | $1.64\times10^{16}$ | $5.3\times10^{15}$ | $2.2\times10^{16}$ |
| DC conductivity (S/m) | $6.3\times10^{7}$ | $5.3\times10^{6}$ | $7.3\times10^{6}$ |
| Transition freq., $\omega_t$(rad/s) | $1.16\times10^{16}$ | $3.8\times10^{15}$ | $1.6\times10^{16}$ |

The dielectric function given in Eqn. (5) is a complex function whose real and imaginary parts are plotted in Figure 1. The figure shows the imaginary part diverging at frequencies below the plasma frequency, whereas the real part becomes the DC conductivity at $\omega = 0$.

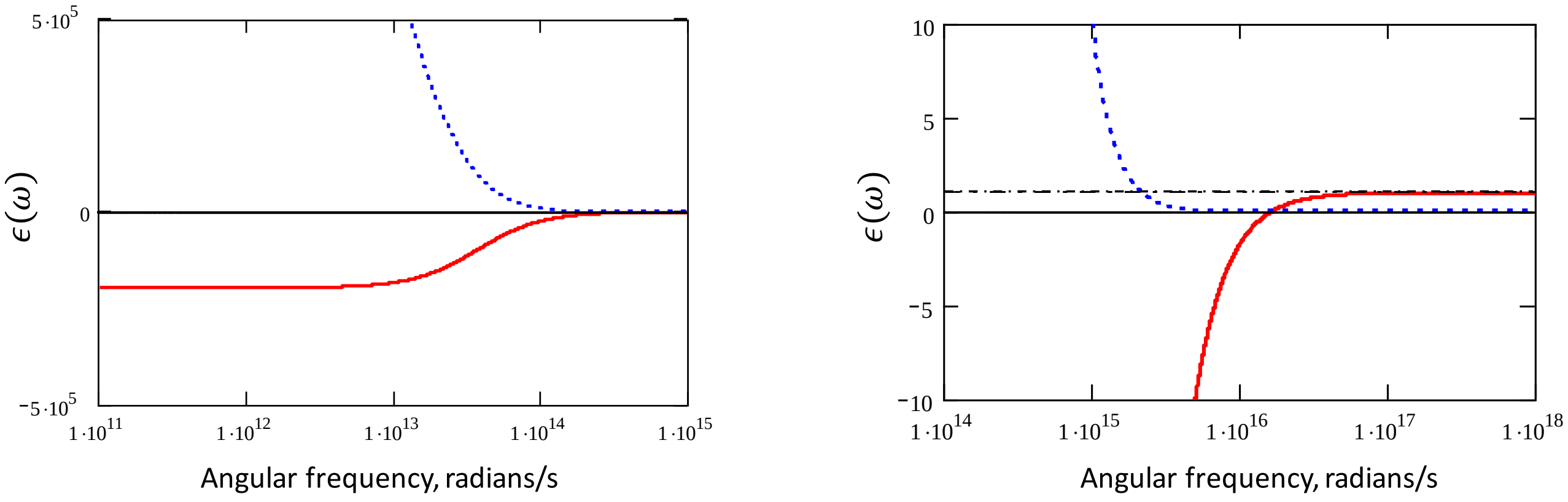

*Figure 1(color): The complex dielectric function for copper at room temperature with different vertical and horizontal scales. The real part is plotted with the solid-red curves and the imaginary part is plotted using dash-blue curves. The plasma frequency is $1.64\times10^{16}$ rad/second and is the frequency where the real part is zero.*

The ratio of the image charge to the real charge (or -$g_c(\omega)$) for copper is shown in Figure 2. The calculations use Eqn. (5) for the cathode dielectric function with copper's plasma frequency and relaxation time. Both the real and imaginary parts diverge at a transition frequency where the real part switches from -1 to 0, and the imaginary part is a narrow, negative peak and is zero everywhere else. This is the frequency where the cathode's image charge transitions from that of a metal (q'/q=-1) to that of a vacuum (q'/q=0). The vacuum-like behavior is identical to the well-known phenomenon of ultraviolet transparency in metals [7]. In the present case, transparency occurs when q'/q=0 and from the beam's perspective the cathode 'disappears'.

The transition between metal and vacuum occurs when

$$\epsilon_c(\omega_t) = -1 \tag{8}$$

where $\omega_t$ is defined as the transition frequency. Setting the real part of the Drude formula for $\epsilon_c$ equal to -1,

$$Re\left(1 + i\frac{\omega_p^2\tau_p}{\omega_t(1-i\omega_t\tau_p)}\right) = -1 \tag{9}$$

and solving for $\omega_t$ gives

$$\omega_t = \sqrt{\frac{\omega_p^2}{2} - \frac{1}{\tau_p{}^2}} \tag{10}$$

From this one sees the transition frequency can be imaginary when the relaxation time is short compared to a plasma period. In this case, image-charge field decays exponentially when

$$\omega_p\tau_p < \sqrt{2}$$

And corresponding, the field oscillates most strongly at the transition frequency, $\omega_t$, when

$$\omega_p\tau_p \geq \sqrt{2}$$

The metasurface described later in Section IVb is designed to have a very short relation time in order to damp the $\omega_t$ oscillations.

Generally, the plasma frequency term dominates making the transition frequency real and equal to

$$\omega_t \cong \frac{\omega_p}{\sqrt{2}} \qquad \text{for} \quad \omega_p\tau_p > \sqrt{2} \tag{11}$$

The transition frequency is thus identical to the surface plasma frequency. Indeed, it is surface plasmon excitations driven by the transient fields of the emitted electron bunch which polarize the cathode surface.

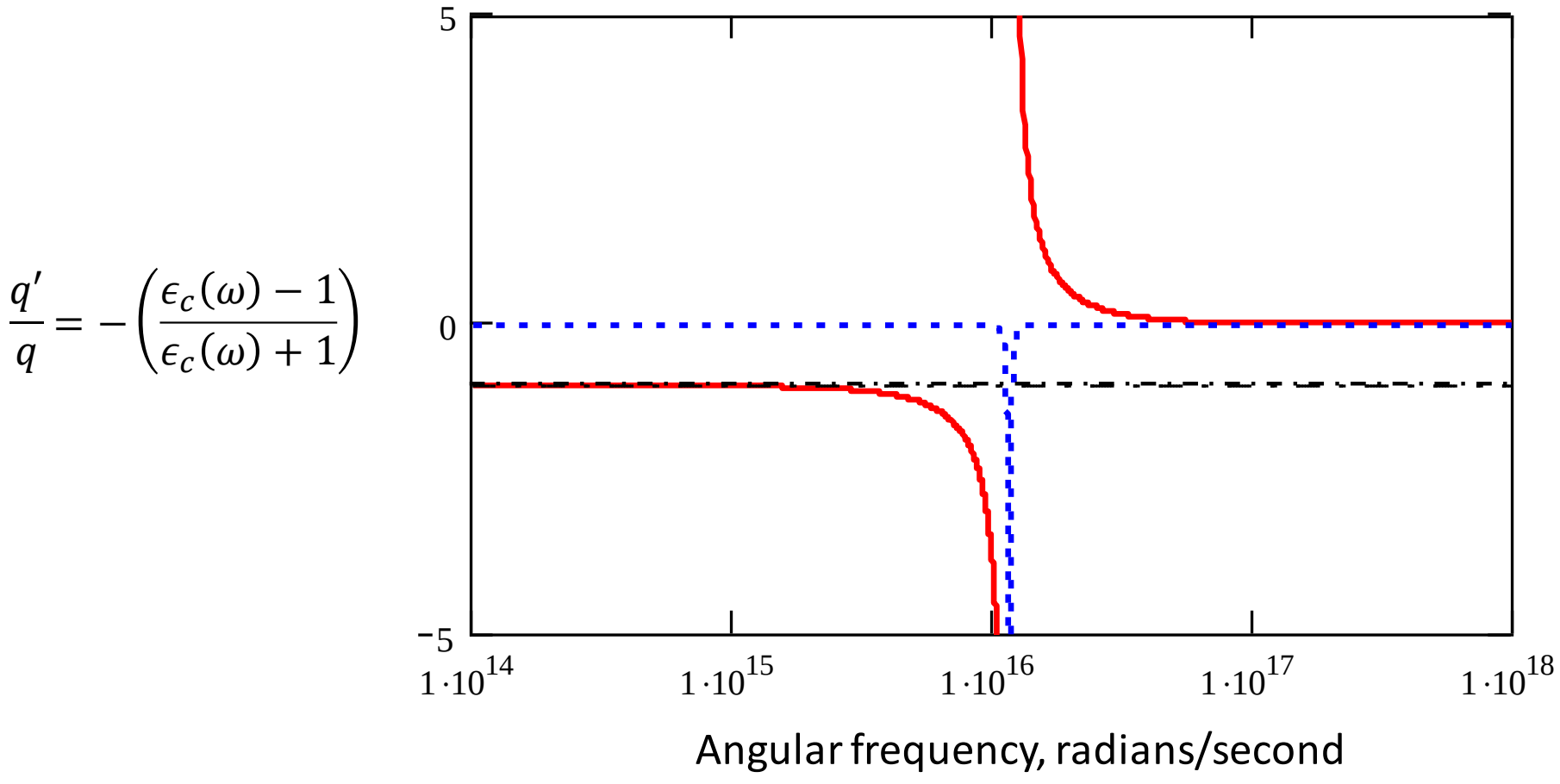

*Figure 2(color): The image-charge to real-charge ratio as a function of frequency, $-g_c(\omega)$, for copper. The dashed blue curve is the imaginary part and the solid red curve is the real part. A dash-dot line is drawn at q'/q=-1 assumed in most current cathode models. The transition frequency for copper is $\omega_t$ =1.16x10$^{16}$ radians/second.*

## III. The image-charge fields in space, time and frequency

This section computes the potentials near the cathode surface for an infinitely thin disk of uniform charge. The analysis derives the transient image-charge electric field experienced by this disk of electrons as it accelerates from the cathode. The Fourier transform of this transient field is multiplied by $g_c(\omega)$ to get the image-charge field as a function of frequency. Fourier transforming this product back to time gives the image-charge field as a function of the disk's time of travel from the cathode.

### a. Potentials at the cathode surface and the Schottky effect

Our model for the electron bunch is an infinitely thin disk of charge with uniform charge density. The results given here can be integrated over a finite bunch length to give the fields for the ideal 'beer can' bunch shape.

The derivation begins with the axial field of a solitary disk of charge as shown in Figure 3. We obtain the image charge field by scaling the disk's charge by q'/q and placing it at the image charge's location behind the cathode. The field from this 'image disk' is then computed at the real disk's z-position on the vacuum side. Assuming constant acceleration in an applied field gives the image-charge field at the disk vs. time of travel from the cathode.

The potential energy surrounding a thin disk has been derived using a Legendre polynomial series [8]. This work gives the axial electric potential at point P the distance $z_P$ from the surface of a disk with radius R and uniform surface charge density $\Sigma_0$ as

$$V_{disk}(z_P) = \frac{\Sigma_0}{2\epsilon_0}\left(\sqrt{{z_P}^2 + R^2} - |z_P|\right) = \frac{Q_0}{2\pi\epsilon_0 R^2}\left(\sqrt{{z_P}^2 + R^2} - |z_P|\right) \tag{12}$$

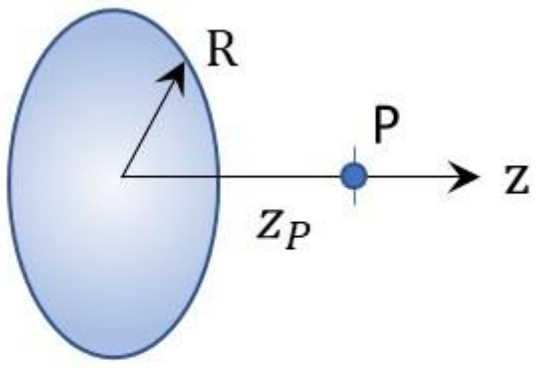

$$\Sigma_0 = \frac{Q_0}{\pi R^2}$$

*Figure 3(color): Parameters of a thin disk of charge.*

The image method places a copy of the real charge at the symmetrical position inside of the cathode to calculate the image-charge field it produces at the disk outside the cathode. Figure 4 shows the image method's configuration for a real-charge disk at $z = z_s$ and its image at $z = -z_s$. The disk and the image charge converge at the cathode surface as $z_s \to 0$. Shifting Eqn. (12) to the image charge location, using the surface loss function for q'/q and evaluating the potential at the disk center, $z_P = 0$, gives the image-charge potential energy at the center of the real-charge disk,

$$V_{total}(z_s) = -E_a z_s - \frac{Q_0}{\epsilon_0 \pi R^2}\left(\frac{\epsilon_c - 1}{\epsilon_c + 1}\right)\left(\sqrt{z_s^2 + \frac{R^2}{4}} - z_s\right) + \phi_{work} \tag{13}$$

The first term is the potential energy of the applied field and the last term is the cathode's work function. The middle term is the image-charge potential energy at the center of a disk with radius $R$ and charge $Q_0$ located $z_s$ from the cathode surface.

Plots of the image-charge plus the applied field potentials are shown in Figure 5. The disk charge is 50-pC disk and the applied field is 50 MV/m. The figure shows the minimum of the potential well disappears for $R > 200$ microns and the peak of the well is approximately 50 microns from the cathode. The image potential is finite at $z_s = 0$ and is equal to

$$V_{image}(z_s = 0) = -\frac{Q_0}{\epsilon_0 2\pi R}\left(\frac{\epsilon_c - 1}{\epsilon_c + 1}\right) \tag{14}$$

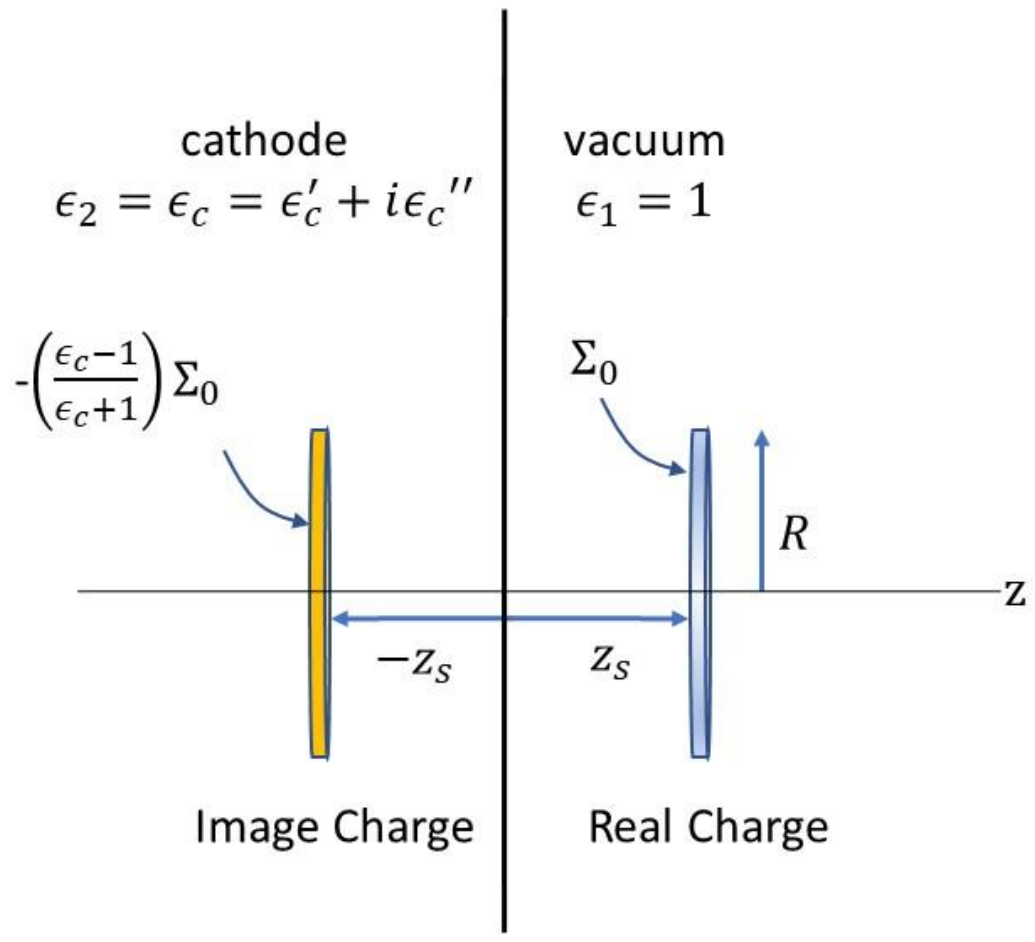

*Figure 4(color): The positions of a real disk of charge and its image with respect to the cathode surface.*

The maxima of the potentials shown in Figure 5 give the Schottky potentials for various radii disks near a metallic cathode. Taking the derivative of Eqn. (13), setting it equal to zero and solving for z gives the potential barrier's maximum location, $z_m$, from the cathode surface. This distance is

$$z_m = \frac{R}{2}\frac{1-\frac{E_a}{E_{disk}}}{\sqrt{1-\left(1-\frac{E_a}{E_{disk}}\right)^2}} \tag{15}$$

where

$$E_{disk} \equiv \frac{Q_0}{\epsilon_0 \pi R^2}\left(\frac{\epsilon_c-1}{\epsilon_c+1}\right) \tag{16}$$

The potential well disappears when the applied field, $E_a$, equals the equivalent image-charge field of the disk, $E_{disk}$. The Schottky potential at the center of the disk is found by evaluating the potential at $z_m$. Doing this gives the Schottky potential for the disk of charge,

$$V_{schottky}(E_a, E_{disk}) = \phi_{disk} = \frac{R}{2}\frac{\frac{E_a}{e}\left(\frac{E_a}{E_{disk}} - 2\right)}{\sqrt{1-\left(1-\frac{E_a}{E_{disk}}\right)^2}} \tag{17}$$

From these relations we can establish the connection between the space-charge limit and the cathode's surface loss function. In photocathode guns a commonly used definition states the SCL is reached when the field of the polarization charge density induced on the surface by the bunch equals the applied field. Equation (15) agrees with this definition since $z_m = 0$ and $\nabla V = 0$ when $E_{disk} = E_a$. This means the SCL field for a disk of charge is frequency-dependent and is proportional to the surface loss function,

$$E_{SCL}(\omega) \equiv \frac{Q_0}{\epsilon_0 \pi R^2}\left(\frac{\epsilon_c-1}{\epsilon_c+1}\right) = \frac{Q_0}{\epsilon_0 \pi R^2}\, g_c(\omega) \tag{18}$$

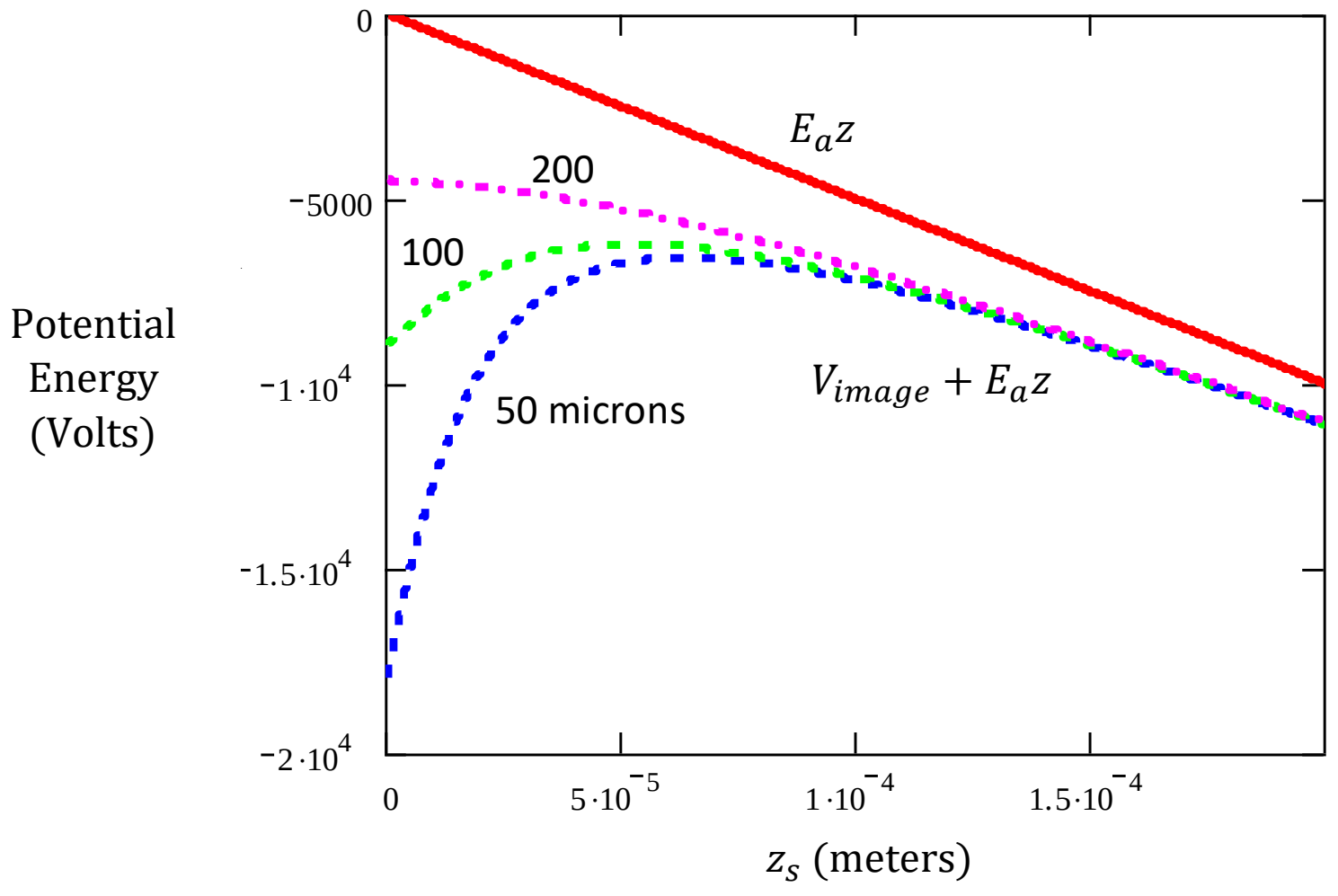

*Figure 5(color): The image-charge potential energy plus the applied potential energy (as given by Eqn. (13)) at the center of a 50-pC disk of charge as a function of its distance from a cathode with q'/q=-1. The potential for an applied field of 50 MV/m is given by the solid-red curve. The dashed curves are the potential energies for disks with 50-, 100- and 200-micron radii. Note that the maximum of the potential barrier is approximately 50 microns from the cathode surface and moves toward the cathode with increasing radius.*

The image-charge potential of a single electron is similarly modified by the cathode's dielectric function. In this case, an electron's total potential energy including the cathode's surface loss function is

$$\phi(z) = \phi_{work} - eE_a z - \frac{e^2}{16\pi\epsilon_0}\left(\frac{\epsilon_c - 1}{\epsilon_c + 1}\right)\frac{1}{z} \tag{19}$$

As described earlier, the Schottky potential is the total potential energy evaluated at its maximum which is

$$\phi_{Schottky} = -\sqrt{\frac{eE_a}{4\pi\epsilon_0}\left(\frac{\epsilon_c - 1}{\epsilon_c + 1}\right)} = -\sqrt{\frac{eE_a}{4\pi\epsilon_0} g_c(\omega)} \tag{20}$$

The relative permittivity of most cathode materials is large compared to 1, therefore $g_c \rightarrow 1$ and Eqn. (20) reverts to the usual expression for the Schottky potential of a single electron. Like the disk model, if $\epsilon_c = 1$ both the image charge and the Schottky potential are zero. This is a disadvantage for metal cathodes which rely on the Schottky effect to lower the effective work function and increase the QE. However, it makes little difference for semiconductor cathodes where the QE is reasonably high.

### b. Time dependence of the image-charge fields

We now calculate of the time-dependent image field at the center of the disk. The potential energy of the image charge at the center of a disk with uniformly distributed charge $Q_0$ a distance $z_s$ from a cathode surface is

$$V_{image}(z_s) = -\frac{Q_0}{\epsilon_0 \pi R^2}\left(\frac{\epsilon_c - 1}{\epsilon_c + 1}\right)\left(\sqrt{z_s^2 + \frac{R^2}{4}} - z_s\right) \tag{21}$$

The divergence of the potential energy, $\vec{E} = -\vec{\nabla} V$, gives us the image-charge's electric field acting on the center of the disk,

$$E_{image}(t) = \frac{Q_0}{\epsilon_0 \pi R^2} \left(\frac{\epsilon_c - 1}{\epsilon_c + 1}\right) \left( \frac{z_s(t)}{\sqrt{z_s(t)^2 + \frac{R^2}{4}}} - 1 \right) \qquad (22)$$

Assuming the disk's center of motion is mostly due to its acceleration in the applied field, the disk's relativistic equation of motion is

$$z_s(t) = \frac{1}{\gamma'} \left( \sqrt{(\gamma' ct)^2 + 1} - 1 \right) \qquad (23)$$

Here $\gamma' \equiv \frac{E_a}{mc^2}$ is the normalized applied field.

Unfortunately, inserting Eqn. (23) into Eqn. (22) and finding a closed expression for the Fourier transform seems impossible. Instead a very good approximation for $E_{image}(t)$ is a heuristic gaussian function. Fitting the width and amplitude of a gaussian to $E_{image}$ gives the following useful and accurate approximation,

$$E_{image}^{gaussian}(t) = -\frac{Q_0}{\epsilon_0 \pi R^2} \left(\frac{\epsilon_c - 1}{\epsilon_c + 1}\right) e^{-\frac{t^2}{2\sigma_t^2}} \qquad t \geq 0 \qquad (24)$$

The rms width of the gaussian, $\sigma_t$, is the image-charge field duration which is a function of the disk radius and the normalized applied field,

$$\sigma_t^2 \equiv \frac{\left(1 + \gamma' R/\sqrt{12}\right)^2 - 1}{(2 \ln 2)\, \gamma'^2 c^2} \qquad (25)$$

Figure 6 shows that $E_{image}^{gaussian}$ closely matches $E_{image}$ which justifies using it to compute the Fourier transforms.

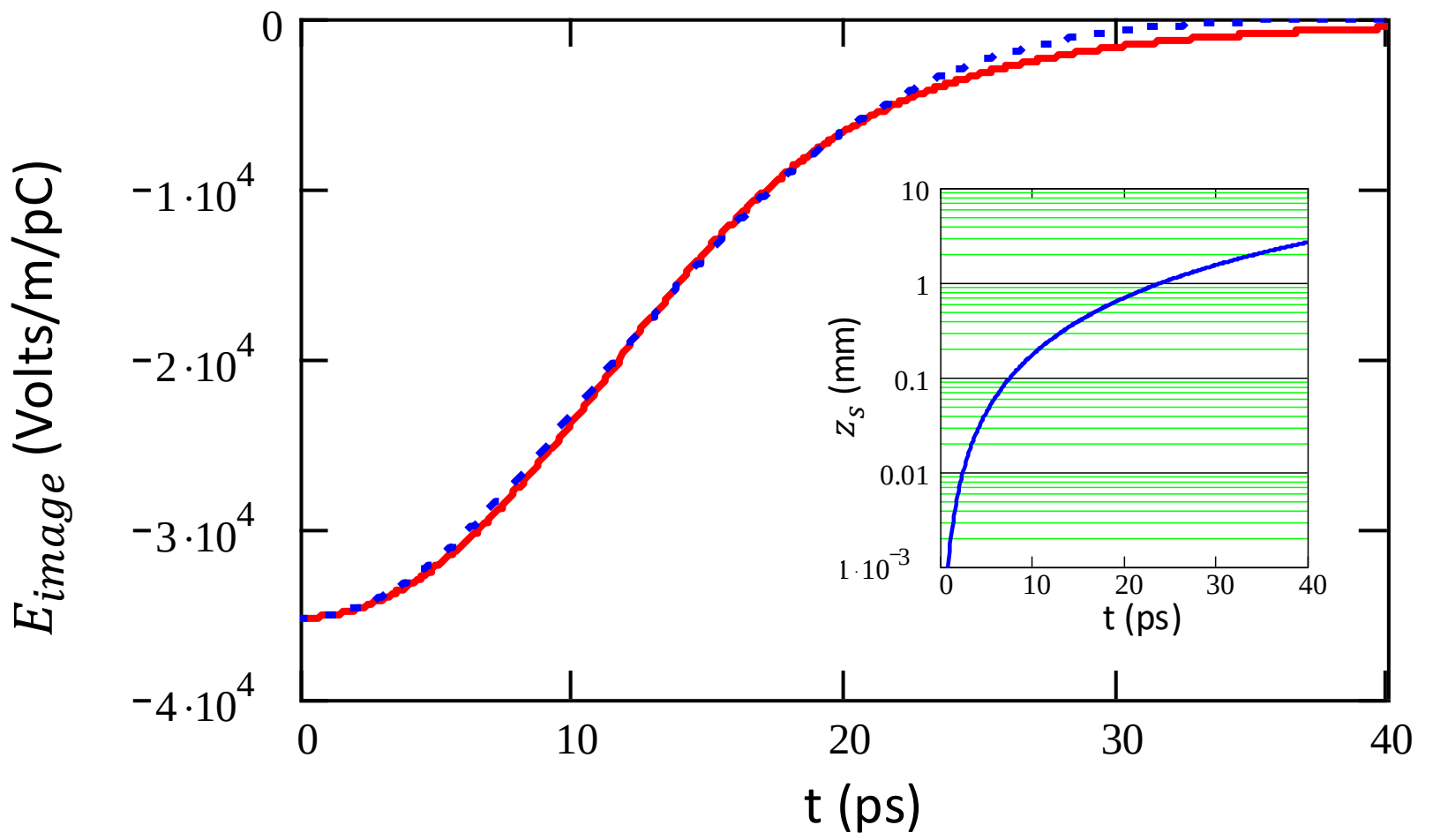

*Figure 6(color): Comparison of the image field given by Eqns. (22) and (23) (solid-red) with the heuristic Gaussian function in Eqn. (24) (dash-blue). The disk radius is 1 mm, $E_a = 20$ MV/m and the cathode dielectric constant is 100. The field is given in units of the field per picocoulomb of negative charge. The insert shows the disk position from the cathode as a function of time.*

Figure 7 shows the image-charge field duration, $\sigma_t$, as a function of the applied field for various disk radii. Clearly the field's duration increases with increasing disk radii and decreases with increasing applied field. Taking the limit of Eqn. (25) as the applied field becomes infinite leads to a minimum value of the image field duration time for a given disk radius,

$$\lim_{\gamma' \to \infty} \sigma_t = \frac{R}{2c\sqrt{\sqrt{3}\ln 2}} \cong 1.5R[mm]\; ps \tag{26}$$

Therefore, even with an infinitely large applied field the electrons will experience the space charge image field for approximately a picosecond unless the cathode radius size is very small.

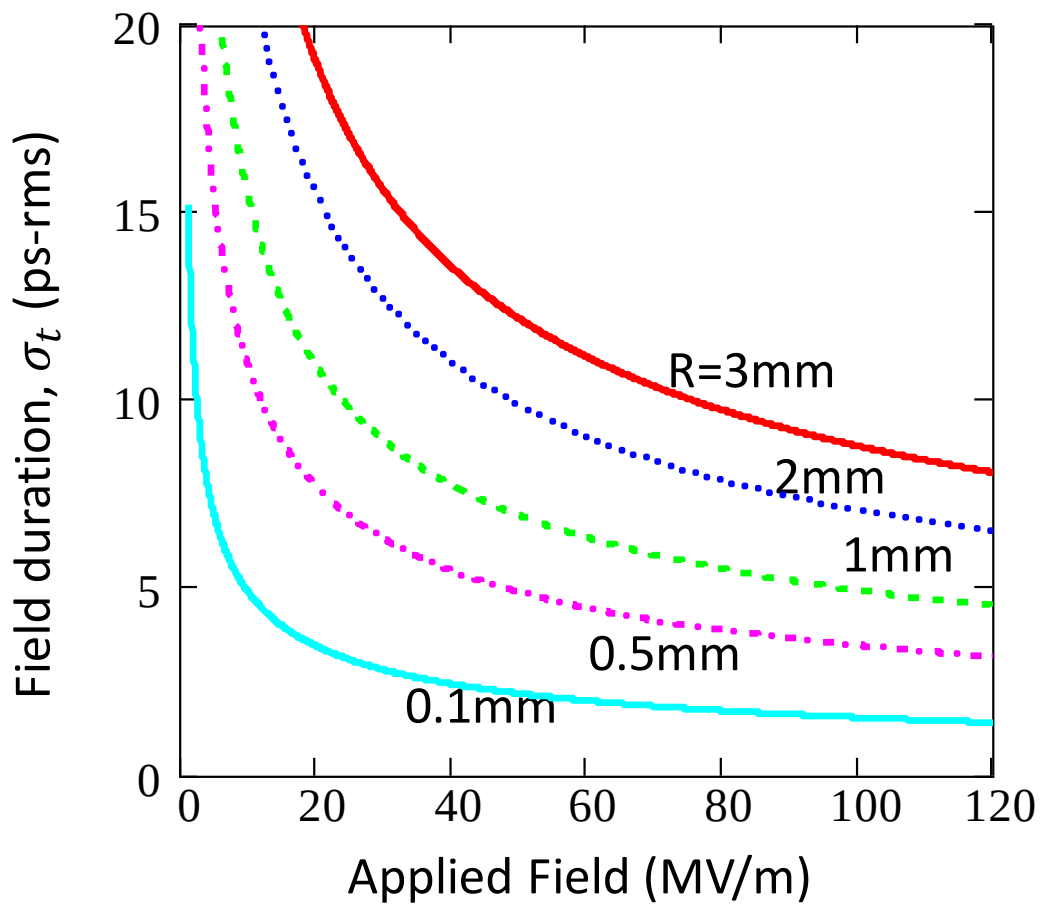

*Figure 7(color): The image field duration time as a function of the applied field for various disk radii.*

### c. Frequency dependence of the image-charge fields

Realizing that the image-charge field is the product of separate frequency- and time-dependent functions, we make these functions explicit by writing the field as,

$$E_{image}^{gaussian}(\omega, t) = \frac{Q_0}{\epsilon_0 \pi R^2}\, g_c(\omega) f(t) \tag{27}$$

Where $g_c(\omega)$ is the cathode's surface loss function and $f(t)$ is defined as the gaussian field shape function,

$$f(t) \equiv e^{-\frac{t^2}{2\sigma_t^2}} \tag{28}$$

The Fourier transform of $f(t)$ is another gaussian given by

$$\tilde{f}(\omega) = \sqrt{\frac{2}{\pi}}\sigma_t e^{-\frac{\sigma_t^2\omega^2}{2}} \tag{29}$$

$1/\sigma_t$ is the rms frequency spread of the transformed image-charge field. The Fourier transform of the image-charge field is thus the product of the surface loss function and the Fourier transform of the field shape function,

$$\tilde{E}_{image}^{gaussian}(\omega) = \frac{Q_0}{\epsilon_0 \pi R^2}\, g_c(\omega)\tilde{f}(\omega) \tag{30}$$

In this product the surface loss function acts like a band-pass filter on $\tilde{f}(\omega)$ allowing only frequencies near the transition frequency to pass. The width of the pass band is related to the relaxation time.

Since $\tilde{E}_{image}(\omega)$ has a gaussian shape centered at $\omega = 0$, it is convenient to define the high frequency edge of $\tilde{E}_{image}(\omega)$ at its half-height as $\omega_e$. Thus, from Eqn. (29),

$$e^{-\frac{\omega_e{}^2\sigma_t^2}{2}} = \frac{1}{2} \tag{31}$$

Solving for the edge frequency, $\omega_e$, gives

$$\omega_e = \frac{\sqrt{2\left|\ln\frac{1}{2}\right|}}{\sigma_t} = 2\left|\ln\frac{1}{2}\right|\sqrt{\frac{\gamma'^2c^2}{\left(1+\gamma' R/\sqrt{12}\right)^2-1}} \tag{32}$$

Like the transition frequency, the edge frequency is a function of the disk radius and the applied field. The temporal behavior and the magnitude of the image-charge field depends upon edge frequency with respect to the transition frequency. Thus, there are three ranges for $\omega_t$ with respect to $\omega_e$ which affect the image-charge field. They are:

If $\omega_t > \omega_e$ then q'/q=-1 with no imaginary part. True for all currently used cathodes.

If $\omega_t = \omega_e$, then q'/q has both real and imaginary parts. The imaginary part is at its maximum. The image-charge field strongly oscillates at the transition frequency which could seed unwanted microbunching instabilities and reduce beam brightness. Or be used to pre-bunch the beam.

If $\omega_t < \omega_e$, then q'/q ~ 0 for both real and imaginary parts. Without the image-charge field, there is no space charge limit to the emission.

### d. The image-charge electric field filtered by the surface loss function

Since $E_{image}(\omega, t)$ is an even function in time, the cosine-representation for the Fourier transform is used [9]. In this paper the Fourier transform of a function is denoted by a tilde symbol. Hence the Fourier transform of the image field is given by

$$\tilde{E}_{image}(\omega) = \frac{2}{\pi}\int_0^\infty E_{image}(\omega, t)\cos\omega t\,dt \tag{33}$$

And its transformation back to the time-coordinate is denoted by the script of the function,

$$\mathcal{E}_{image}(t) \equiv \int_0^\infty \tilde{E}_{image}(\omega)\cos\omega t\,d\omega \tag{34}$$

$\tilde{E}_{image}(\omega)$ is the Fourier spectrum of the image-charge field seen by the disk as it accelerates from the cathode. Assuming the gaussian formulation for $E_{image}(t)$ is valid, the Fourier spectrum is easily computed,

$$\tilde{E}_{image}^{gaussian}(\omega) = -\sqrt{\frac{2}{\pi}}\frac{Q_0}{\epsilon_0\pi R^2}\left(\frac{\epsilon_c(\omega)-1}{\epsilon_c(\omega)+1}\right)\sigma_t e^{-\frac{\omega^2\sigma_t^2}{2}} \tag{35}$$

Equation (35) is the Fourier frequency spectrum of the image-charge field responding to the electrical impulse of the accelerating disk. Transforming $\tilde{E}_{image}^{gaussian}(\omega)$ back to time gives a modified time-dependent image-charge field which now includes the cathode's response,

$$\mathcal{E}_{image}^{gaussian}(t) = -\sqrt{\frac{2}{\pi}}\frac{Q_0}{\epsilon_0\pi R^2}\sigma_t\int_0^\infty\left(\frac{\epsilon_c(\omega)-1}{\epsilon_c(\omega)+1}\right)e^{-\frac{\omega^2\sigma_t^2}{2}}\cos\omega t\,d\omega \tag{36}$$

Using Drude's theory for the dielectric function (see Eqn. (5)), the integral can be written in terms of the plasma frequency and the relaxation time of the cathode surface,

$$\mathcal{E}_{image}^{gaussian}(t) = -\sqrt{\frac{2}{\pi}}\frac{Q_0}{\epsilon_0\pi R^2}\sigma_t\int_0^\infty\left(\frac{\omega_p^2}{\omega_p^2-2\omega\left(\omega+\frac{i}{\tau_p}\right)}\right)e^{-\frac{\omega^2\sigma_t^2}{2}}\cos\omega t\,d\omega \tag{37}$$

For cathode surfaces with high plasma frequencies, then $\epsilon_c \gg 1$ therefore $\left(\frac{\epsilon_c-1}{\epsilon_c+1}\right) \to 1$ and q'/q →-1. In this case, the Fourier transform of $\tilde{E}(\omega)$ returns the original image field such that $\mathcal{E}_{image}(t) = E_{image}(t)$. This is not a very interesting result. However, if one instead uses a frequency-dependent dielectric function with a transition frequency low enough to overlap with the field's Fourier spectrum then some interesting effects can be discovered as described in Section IV.

### e. Transverse size of the polarization surface charge density

In this subsection the transverse size of the cathode needed for the metasurface properties to dominate is determined. The metasurface needs to be large enough to include all the polarization charge induced on the cathode's surface by the disk of charge. The polarization charge density is derived using the image method. The distribution of the polarization charge determines the areal size of the cathode's metasurface.

The polarization charge density on the cathode surface at radius $\rho$ induced by a charge q at a distance $d$ from a surface is [4]

$$\sigma_{pol} = -\frac{q}{2\pi}\left(\frac{\epsilon_c-1}{\epsilon_c+1}\right)\left[\frac{d}{(\rho^2+d^2)^{\frac{3}{2}}}\right] \tag{38}$$

The total polarization charge density at the point $(\rho, \alpha)$ on the cathode is found by integrating over the disk coordinates $(r, \alpha')$,

$$\sigma_{pol}(\rho, \alpha) = -\frac{d}{2\pi}\left(\frac{\epsilon_c-1}{\epsilon_c+1}\right)\int_0^{2\pi}\int_0^R \frac{\Sigma_0 r dr d\alpha\prime}{[\rho^2+d^2+r^2-2\rho r\cos(\alpha-\alpha\prime)]^{3/2}} \tag{39}$$

Figure 8 illustrates the three views of the cathode-disk geometry. These drawings show an infinitesimal of charge, $dQ = \Sigma_0\, r dr\, d\alpha'$, located on the disk at coordinates $(r, \alpha')$ polarizing charge on the cathode surface at $(\rho, \alpha)$. The bracketed quantity in the integrand's denominator is the square of the 3-dimensional distance between the points $(\rho, \alpha)$ on the cathode and $(r, \alpha')$ on the disk (Fig. 8, side view).

Since we're interested in the radial extent of the polarization charge and because of the disk's axial symmetry, there is no loss of generality setting $\alpha = 0$ and determining the charge density profile along the x-axis. Doing this and performing the radial integral gives

$$\sigma_{pol}(\rho, \alpha) = -\frac{d}{2\pi}\left(\frac{\epsilon_c-1}{\epsilon_c+1}\right)\Sigma_0\int_0^{2\pi}\left\{-\frac{s^2-\rho R\cos\alpha'}{(R^2+s^2-2\rho R\cos\alpha')^{\frac{1}{2}}}+s\right\}\frac{d\alpha\prime}{s^2-\rho^2\, cos^2\alpha\prime} \tag{40}$$

where $s^2 = \rho^2 + d^2$. This expression is numerically integrated to obtain the polarization charge density.

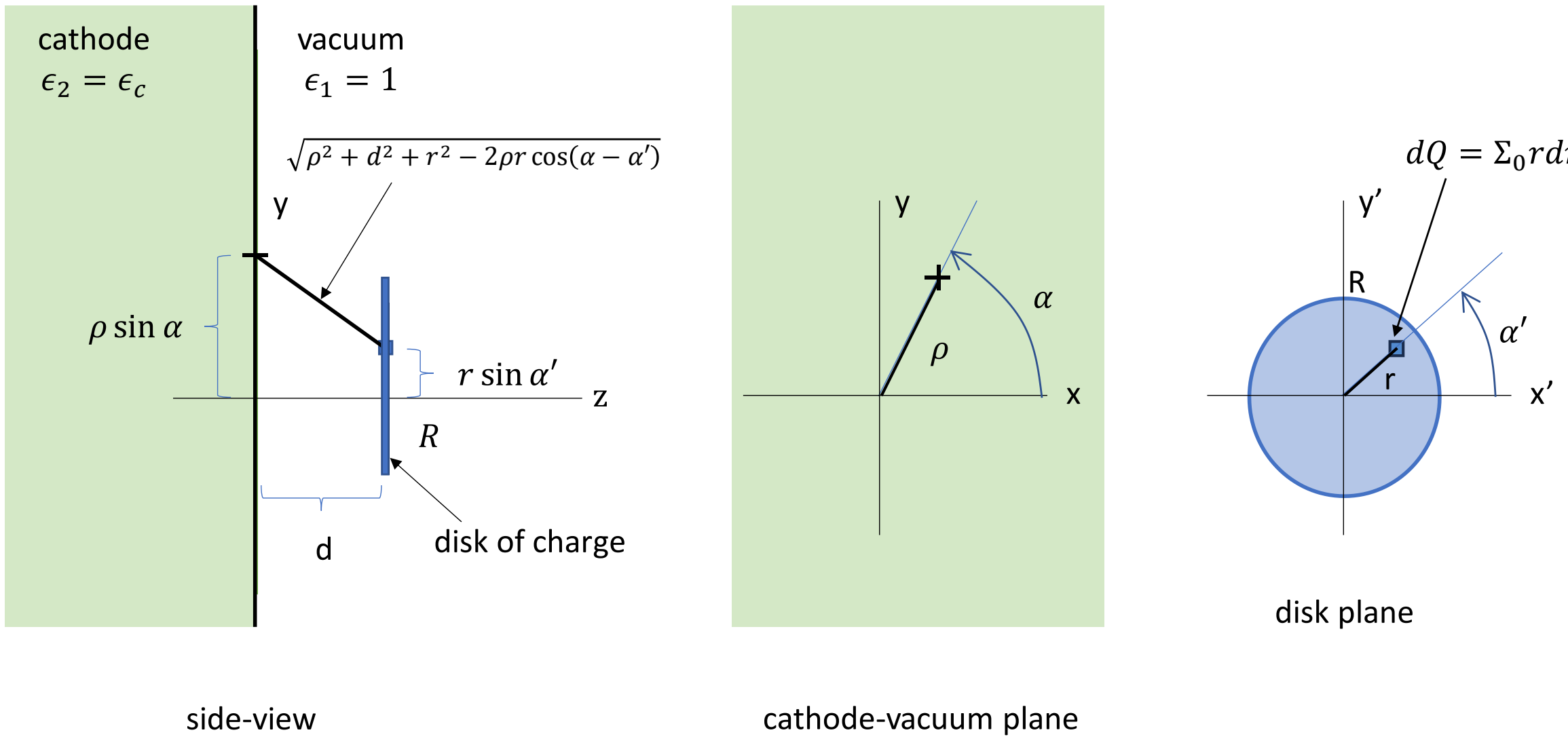

*Figure 8(color): The geometry for calculating the polarization charge density on the cathode surface. The drawings show the distance between the point $(\rho, \alpha)$ on the cathode surface and the point $(r, \alpha')$ on the disk. The disk has a uniform surface charge density of $\Sigma_0$. The polarization charge density at the point on the cathode is given by integrating the disk's surface charge density over r and $\alpha'$. The x-axis is into the page in the side view and the z-axis is out of the page in the cathode-vacuum and disk plane views.*

Figure 9 shows the polarization charge density along the x-axis for a 100-pC, 1-mm radius disk positioned 0.03, 0.3 and 3-mm from the cathode. The figure indicates the transverse extent of the polarization charge is about the same as the disk size which in this case is 2-mm in diameter. From this transverse distribution, we conclude that the metasurface should be at least twice the beam size at the cathode to include tails of the distribution.

The thickness of the polarized layer is only a few angstroms [10]. Which would imply that the meta-cathode can be made very thin. However, the actual thickness will depend upon the metasurface design and its surface loss function. This is discussed with more detail in Section IVb.

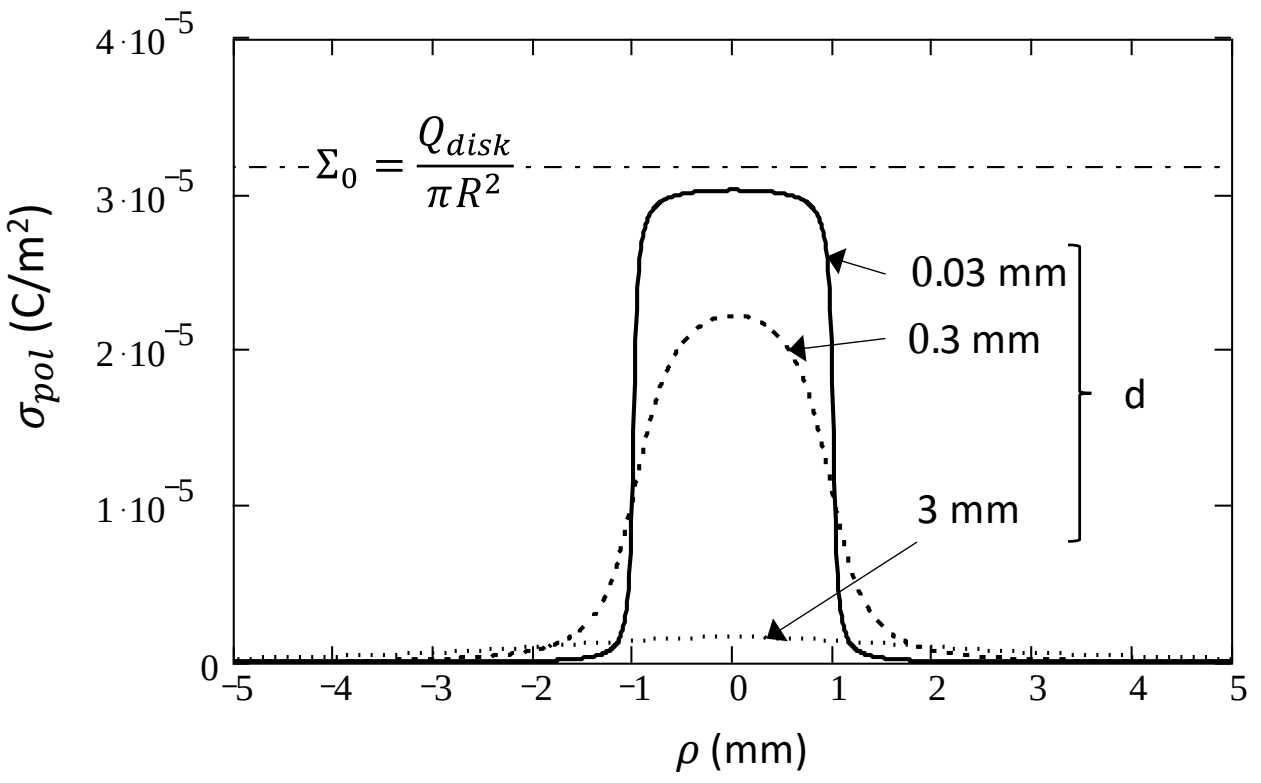

*Figure 9: The polarization charge density on the cathode surface for a R=1 mm radius disk with $Q_{disk}$ = 100 pC. Line-outs for the surface charge density are shown for the disk at d=0.03, 0.3 and 3 mm from the cathode. The surface charge density of the disk, $\Sigma_0$, is shown with a dash-dot line.*

## IV. Cathode design with metasurfaces

A metasurface is a surface composed of a pattern of structures each of which is smaller than the wavelength of the field of interest. At wavelengths longer than their spacing, these structures interact with radiation like atoms in an atomic lattice. Such an artificial 'atomic' structure can have significantly different electronic properties than those of a natural bulk material. Varying the type of structures and their patterns produces a wide range of electrical and magnetic properties including many which are otherwise unobtainable in nature. In the case of cathode design, an important electronic property is the dielectric function which affects the electron dynamics via the image charge also known as the surface loss function. The dielectric function is determined by the structure of the metasurface and, as will be shown, the metasurface can be structured to mitigate the cathode's SCL.

This section consists of four subsections. The first subsection describes a wire-array metasurface whose plasma frequency is easily varied to study the image-charge electric field. The second subsection describes a proposed meta-cathode design which reduces the SCL field. The third subsection descrbes incorporating electron emitters into the metasurface to complete the holistic meta-cathode design. And the final, fourth subsection discusses the practical issues of putting an array of fine wires in a high-field electron gun.

### a. The surface loss function of a wire-array metasurface

Studies of the electrical properties of 'artificial dielectrics' began in the 1940's. An early description of wires arrays in terms of a plasma was given by Rotman who simulated microwave antennas with 1D, 2D and 3D wire arrays [11]. More recently wire arrays which mimic materials with very low plasma frequency are discussed by Pendry et al. [12]. In his work, Pendry describes a 3D cubic lattice of wires and emphasizes the importance of the wire radius being much smaller than the wire spacing.

Here we study a 1D metamaterial formed from a rectangular array of thin metal wires aligned parallel to the z-axis. The direction of the wires determines the electric field polarization coupling to the array's dielectric function. Other polarizations are unaffected. In Fig. 10 the wires are aligned along the z-axis, therefore only the z-component of the electric field is affected by the array. If the wires are thin compared to the wire spacing, the array is nearly transparent to the x- and y-components of the applied field. Making a 3D wire array would couple the array to all three components of the field. Each component of the diagonal dielectric function tensor would be determined by the wire spacing, radius and conductivity of the 1D subarray aligned along that component's axis.

The 1D metamaterial shown in Fig. 10 has a plasma frequency with z-polarization given by [13]

$$\omega_{wire}^2 = \frac{2\pi c^2}{a^2 \ln\frac{a}{r}} \tag{41}$$

and relaxation time

$$\tau_{wire} = \frac{r^2 \sigma_{wire}}{2\epsilon_0 c^2} \ln\frac{a}{r} \tag{42}$$

Here $a$ is the center-to-center distance between the wires, $r$ is the wire radius, and c is the vacuum speed of light. The conductivity of each wire is $\sigma_{wire}$. The host material between the wires is assumed to be vacuum. Applying the theory described in Section II, values for the three array parameters ($a$, $r$ and $\sigma_{wire}$) can be found which reduce the transition frequency by six-orders of magnitude from $10^{16}$ to $10^{10}$ radians/second.

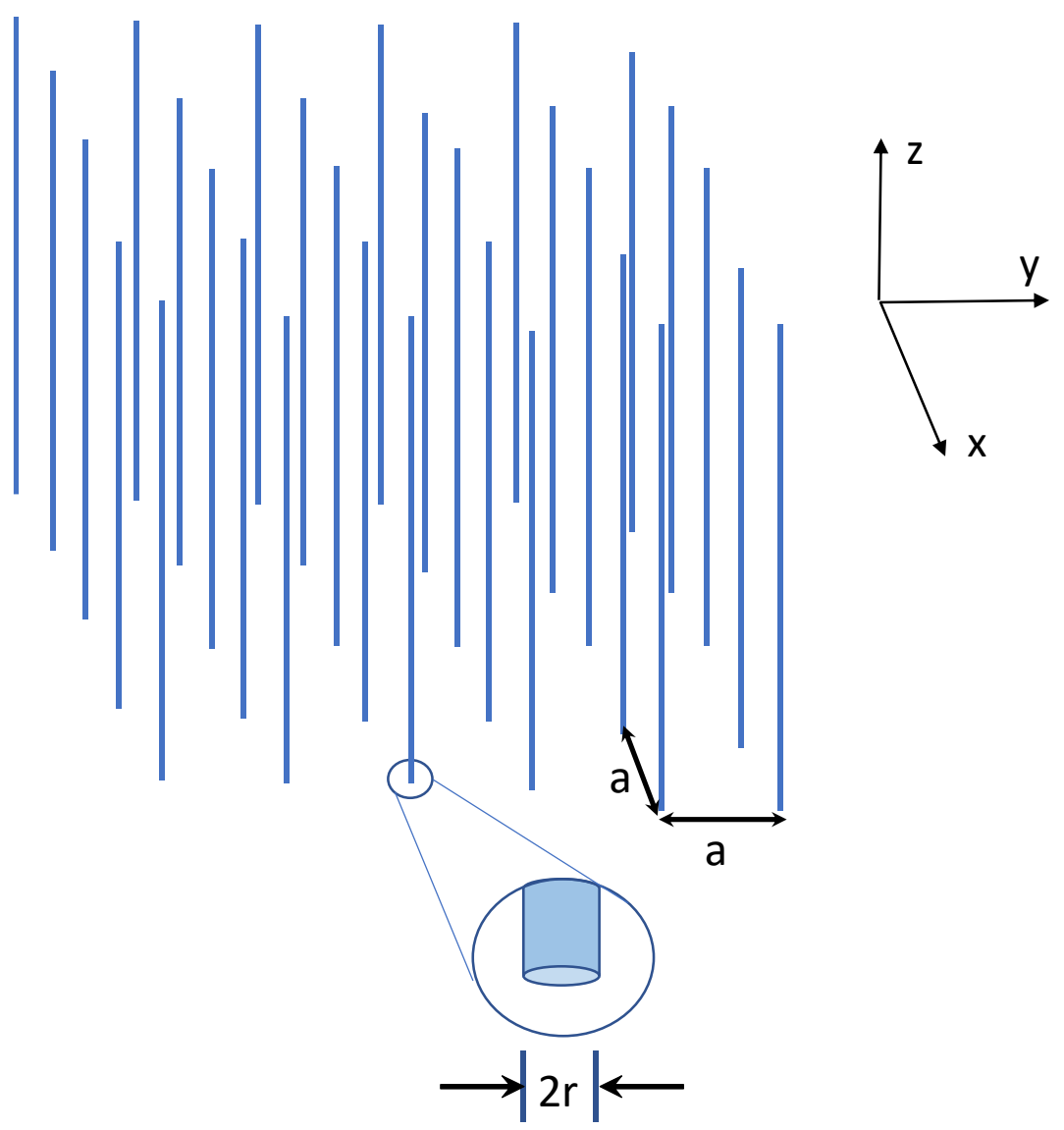

*Figure 10(color): A 1D wire-array metamaterial consisting of a rectangular array of parallel wires with spacing a and wire radius r. These dimensions combined with the wire conductivity $\sigma_{wire}$ specify the plasma frequency and the relaxation time of the metamaterial. Since the wires are aligned along the z-axis, it is only the z-component of the electric field which can electrically polarize the wires. The x- and y-field components are unaffected by the array.*

Table II gives the characteristics of metasurfaces for three wire spacings corresponding to the three frequency regions described earlier. Figure 11 shows the image-charge fields and $g_c(\omega)$. For all three spacings, the wire radius is 1-micron and the wire conductivity is that of copper. The applied field is 50 MV/m which combined with a $R$ = 1 mm disk radius gives $\sigma_t$ = 6.9 ps-rms. The edge frequency is $\omega_e$=1.7x10$^{11}$ radians/second. The figure shows wire spacings a=0.5, 1.3 and 10 mm correspond to $\omega_t > \omega_e$, $\omega_t \cong \omega_e$ and $\omega_t < \omega_e$, respectively. The image-charge fields shown in Fig. 11 confirm the behavior described in Section IIIc for the three frequency ranges.

*Table II. Parameters of rectangular array of vertical wires for three wire spacings. The wire conductivity is $\sigma_{wire}$= 6.3x10$^7$ S/m and the wire radius is r = 1 micron. The field transit time is $\sigma_t$ = 6.9 ps. The edge frequency is $\omega_e$=*1.7x10$^{11}$ radians/second

| frequency range | $\omega_t > \omega_e$ | $\omega_t \cong \omega_e$ | $\omega_t < \omega_e$ |
|---|---|---|---|
| wire spacing, a (mm) | 0.5 | 1.3 | 10 |
| array plasma freq., $\omega_{wire}$ (radians/sec) | 6.0x10$^{11}$ | 2.2x10$^{11}$ | 2.5x10$^{10}$ |
| relaxation time, $\tau_{wire}$ (sec) | 2.5x10$^{-10}$ | 2.8x10$^{-10}$ | 3.6x10$^{-10}$ |
| transition frequency, $\omega_t$ (radians/sec) | 4.3x10$^{11}$ | 1.5x10$^{11}$ | 1.7x10$^{10}$ |

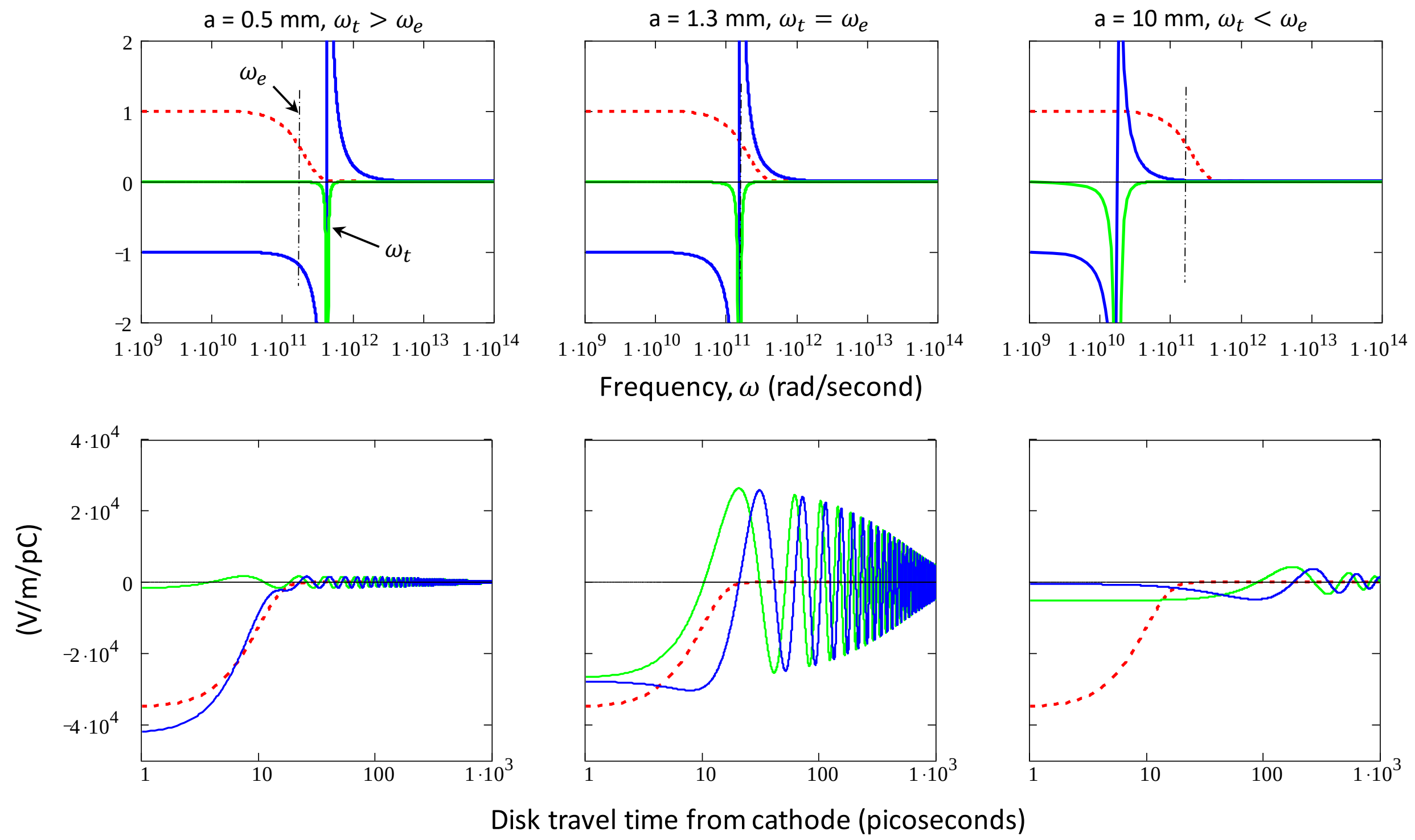

*Figure 11(color): The Fourier spectra and the image-charge fields vs. time for three wire array metamaterials with wire spacings of a=0.5, 1.3 and 10 mm corresponding* to $\omega_t > \omega_e$, $\omega_t \cong \omega_e$ and $\omega_t < \omega_e$, respectively. *The other array characteristics are given in Table II. The applied field is not shown. Upper plots: The image-charge field frequency spectrum normalized at* $\omega = 0$, *and the real and imaginary parts of the surface loss function. The vertical dash-dot line is the edge frequency at* $\omega_e$*=1.7x10*$^{11}$ radians/second. *Red-dash:* $\tilde{f}(\omega)/\tilde{f}(0)$. *Blue-solid:* $= -Re\ g_c(\omega)$. *Green-solid:* $-Im\ g_c(\omega)$.
*Lower plots: The image-charge field at the disk center. Blue-solid:* $Re\ \mathcal{E}_{image}^{gaussian}(t)$*;*
*Green-solid:* $Im\ \mathcal{E}_{image}^{gaussian}(t)$*; Red-dash:* $E_{image}(t)$ *with q'/q= -1.*

### b. Suppressing the SCL using a wire-array meta-cathode

As discussed earlier, the metasurface should be approximately two-times the disk size. In addition, there should be perhaps hundreds of meta-atoms across the disk's diameter. Given a typical beam diameter of 1 mm then says the meta-atom spacing should be ~ 1mm/100 or $a \leq 10$ microns for the wire-array.

The length of the wires can be estimated from the relaxation time. Although the model assumes the wires are isolated and suspended in space, this condition is clearly violated since in our meta-cathode one end is connected to the cathode surface. However, the assumption can be made valid again provided the wires are long enough that a signal sent from the free end decays before reaching the attached end. Since the fastest any signal can travel is the speed of light and the signal decays with the relaxation time, a reasonable estimate is the wire needs to be longer than $c\ \tau_{wire}$ (Figure 12).

For the wire parameters given in Table II, $c\ \tau_{wire}$ ranges from 2.7 to 8.1 cm. Meta-cathodes with this geometry would be a challenge to fabricate and lead to an impractical cathode size for most electron guns. Therefore, we need to search for more realistic array parameters.

Equation (44) shows the relaxation time depends quadratically upon the wire radius and linearly upon the wire conductivity but only logarithmically upon the wire spacing. Therefore, the relaxation time can be reduced orders of magnitude by using small radius wires with low conductivity. The low conductivity also damps the oscillations at $\omega_t$.

With these considerations in mind, a meta-cathode consisting of a rectangular wire array (1D Rotman metamaterial) standing normal to a conducting substrate is proposed. Based upon calculations shown below, the wire radius is 10 nm and the wire spacing is 10 microns. This design easily meets Pendry's thinness requirement since $a/r$ = 1000. In addition, the spacing corresponds to 100 meta-atoms across a 1-mm diameter area of photoemission which satisfies the several meta-atoms requirement. The wire conductivity is only 0.001-times copper's conductivity to damp the $\omega_t$-oscillations of the image-charge by reducing the relaxation time. This leads to fraction of a femtosecond relaxation times, imaginary transition frequencies, and a few nanometer minimum wire lengths as shown in Table III.

*Table III. Parameters for rectangular array of vertical wires. In all cases, the wire conductivity is 1/1000 that of copper ($\sigma_{wire}$ = 6.4x10$^4$/$\Omega/m$) and the wire radius is r=10 nm. The normalized applied field is $\gamma'$ = 98 ($E_a$= 50 MV/m) and the image field duration time is $\sigma_t$=6.9 ps-rms. These parameters are used to compute the curves in Fig. 13.*

| frequency range | $\omega_t > \omega_e$ | $\omega_t \cong \omega_e$ | $\omega_t < \omega_e$ |
|---|---|---|---|
| wire spacing, a (microns) | 1 | 2.5 | 10 |
| array plasma freq., $\omega_{wire}$ (radians/sec) | 3.5x10$^{14}$ | 1.3 x10$^{14}$ | 2.9 x10$^{13}$ |
| relaxation time, $\tau_{wire}$ (fs) | 0.018 | 0.022 | 0.027 |
| min. wire length, $c\tau_{wire}$ (nm) | 4.6 | 6.6 | 8.2 |
| transition frequency, $\omega_t$ (radians/sec) | 5.5i x10$^{16}$ | 4.6i x10$^{16}$ | 3.7i x10$^{16}$ |

A schematic drawing of a wire for the proposed meta-cathode design is shown in Fig. 12. The wire parameters are given by the $a$=10-microns column in Table III. Each wire attaches to the substrate through a ~10-nm thick insulator. The insulator is there to electrically isolate the wires from each other to preserve the assumption of perfect electrical isolation between the wires. The insulating layer should be very thin in order to be "transparent" to ~1 GHz RF fields and absorb little RF power. It should be uniformly thick across the entire area of emission.

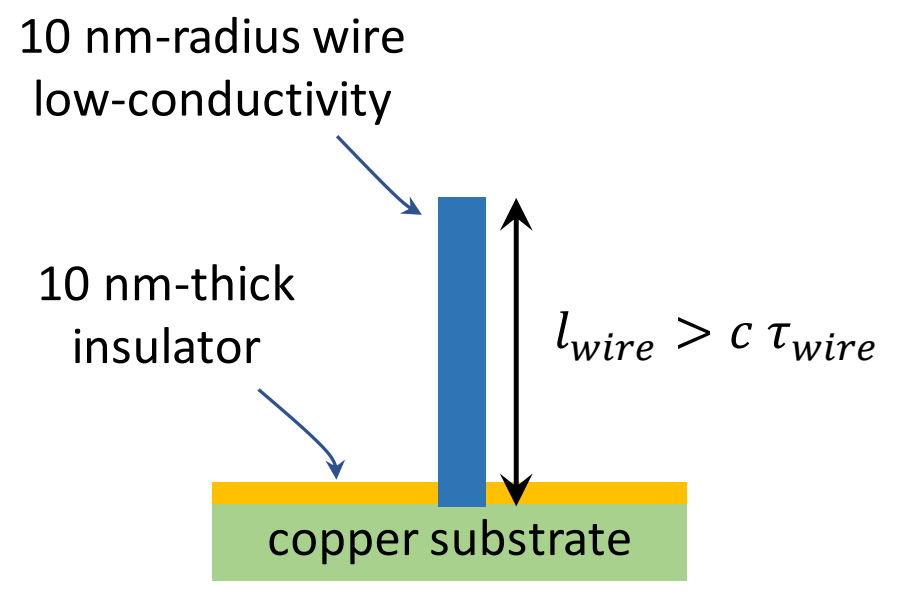

*Figure 12(color): The relaxation time determines the minimum wire length and the low-conductivity damps the $\omega_t$-oscillations. The 10-nm insulator blocks surface plasmons travelling between wires.*

This configuration is like that of a topological insulator where metal conductors are on top of a bulk insulator and the insulator is 'protecting' wires from the surface's low plasma frequency. But in this case the bulk insulator is only 10-nm thick. The thin insulator blocks surface plasmons from travelling between the wires to increase their electrical isolation. The wires penetrate the insulator and connect with the copper substrate which provides each wire with current and cooling.

As in the previous analysis, the wire spacing is varied to put $\omega_t$ above, near and below $\omega_e$. Table III lists the metasurface parameters for these three cases. The transition frequencies are imaginary indicating the image fields quickly decay as they propagate on the surface and there are no oscillations since the transition frequency is imaginary. The very short relaxation times of tens of attoseconds give a very small minimum wire length of $c\tau_{wire}$ ~8 nm.

Figure 13 shows the real and imaginary parts of the surface loss functions and image-charge fields vs. frequency and time for the three wire spacings. At $a$ = 1-micron spacing, we have $\omega_t > \omega_e$ and the surface behaves like the usual cathode with q'/q = -1 and a little imaginary part. The lower plot shows the real part of the field is identical to the nominal, q'/q=-1 image-charge field and the imaginary field is small. At $a$ = 2.5-microns, $\omega_t = \omega_e$ but the $\omega_t$-oscillations are strongly damped by the 0.022 fs relaxation time. At $a$ = 10-microns, $\omega_t < \omega_e$ and q'/q ~ 0, which we choose for the meta-cathode design since the image-charge field is minimal.

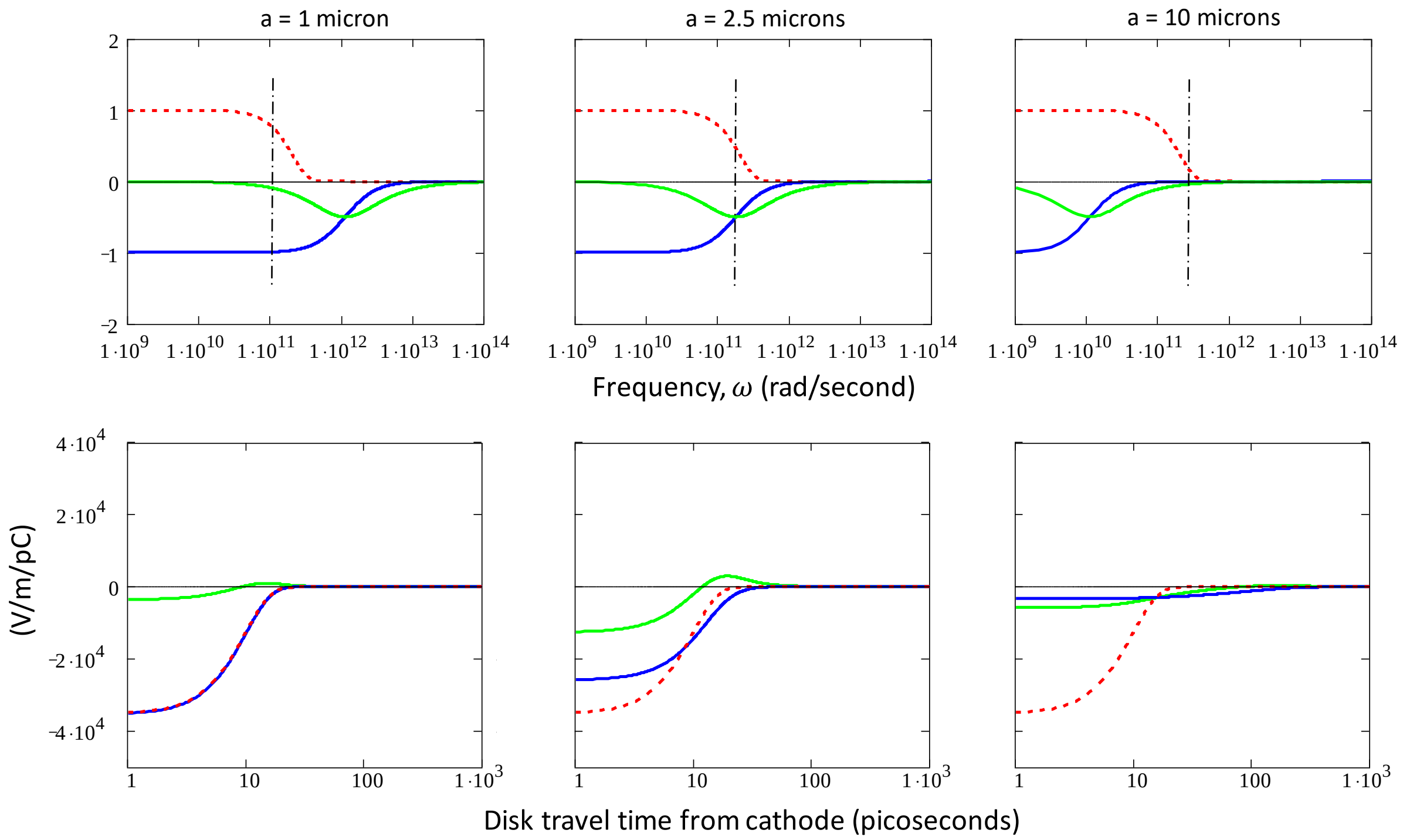

*Figure 13(color): The Fourier spectra and image-charge fields for metasurfaces with wire spacings of a=1.0, 2.5 and 10 microns corresponding to transition frequencies above, equal to, and below* $\omega_e$*, respectively.*

*Upper plots: The image-charge field frequency spectrum normalized at ω = 0 and the real and imaginary parts of the surface loss function. The vertical dash-dot line is at* $\omega_e$*=1.7x10*$^{11}$ radians/second.

*Red-dash:* $\tilde{f}(\omega)/\tilde{f}(0)$. *Blue-solid:* $-Re\ g_c(\omega)$ *Green-solid:* $-Im\ g_c(\omega)$.

*Lower plots: The image-charge field at the disk center. Blue-solid:* $Re\ \mathcal{E}_{image}^{gaussian}(t)$;

*Green-solid:* $Im\ \mathcal{E}_{image}^{gaussian}(t)$; *Red-dash:* $E_{image}(t)$ *with q'/q= -1.*

### c. Holistic wire-array cathode design

In a holistic cathode design, the geometry of electron emission is combined with the applied and space-charge fields at the cathode surface to define the complete meta-cathode. This involves including realistic sources of electron emission to the metasurface design shown in Fig. 12.

Electron emission can be from two types of areas of the wire-array meta-cathode. One is from the tips of the wires and the other from the region between the wires. This sub-section describes meta-cathode concepts for these emission geometries.

The work function of the emitting tips can be reduced by 'decorating' them with diamondoid monolayers [14]. Making the emitting tips longer than the surrounding tips would give them higher fields due to field enhancement. In fact, the elevation between the tips could be adjusted to finely tune the relative potential energies of different beamlets at the micron-scale and add features such as a radial correlation with energy. Afterall for an applied field of $E_a$= 50 MV/m, a difference in pillar height of 10-nm corresponds to a 0.5-volt energy difference.

Electron emitters placed between the wires should be small and positioned halfway between the wires to minimize and symmetrize interactions. In this case, each emitter's height should be much less than the heights of the surrounding wires, so as not to interfere with the dielectric properties of the metasurface.

And finally, it is worth noting that a 10-nm wire radius is about the size of a quantum dot, and Q-dots are already being proposed as cathodes for electron sources. Therefore, mounting Q-dots on the top of a few hundred nm tall wires spaced every 10-microns would be a natural integration of the two technologies. The conductivity of the wires would still need to be in the $10^4$ S/m range.

### d. Practical considerations

Figure 13 illustrates an important feature of the proposed meta-cathode design at a=10 microns. In particular, the surface loss function becomes 1 (q'/q=-1) below approximately 1 GHz. Thus, at low RF frequencies the surface responses like that of a metal. This is because long wavelength RF cannot resolve the very thin wires and the surface appears as a uniform metallic surface, albeit with slightly reduced surface conductivity. The low frequencies mostly 'see' the surface below the wire array where the conductivity is 1000-times better and there are few RF losses.

However, there remains the very important issue of electrical breakdown of the meta-surface with a high applied field. This is potentially a serious technical problem because the low conductivity wires could experience excessive resistive heating by the current drawn through each thin wire. And possibly burnup if they draw too much current during a breakdown while RF or HV processing the gun. Another concern is multipacting which may occur between the wires and other surfaces. These and other effects at the nanometer to micron scales deserve further theoretical and experimental study.

## V. Conclusions

This paper shows it is possible to design cathode surfaces which mitigate the space-charge limit due to the image charge, and make the cathode appear 'transparent' to the electrons. The resulting increase in the total accelerating electric field at the cathode's surface can be significant for high-charge bunches. For example, Fig. 13 shows the image-charge field at the surface of the typical cathode with 1-mm radius is $-3.5\times10^4$ V/m/pC. Thus, the SCL-field for a 100-pC bunch is -3.5 MV/m. This is nearly equal to the 4.4 MV/m applied field of the Cornell DC gun [15] [16], leaving a cathode surface field of only 0.9 MV/m. Instead, the meta-cathode design with the 10-micron wire spacing and wire radius of 10-nm has an image field of -2x10$^3$ V/m/pC which is 17-times smaller. Therefore, the meta-cathode's SCL-field at 100-pC is only 200 KV/m leading to a 4.6x higher cathode field of 4.2 MV/m.

The ability to accurately fabricate large areas of nanometer scale wires, pillars, columns and cones is essential to making meta-cathodes a reality. A meta-cathode will require precisely placing 10$^4$ or more meta-atoms on a rather sparse pattern over a relatively large diameter of a few mm. Fortunately, recent research of a technique called nanomolding [17] [18] shows such precise fabrication is indeed possible. Other work suggests filling the 10-micron space between the wires with nanometer sized electronic circuits comprised of nanoinductors, nanocapacitors and nanoresistors is also possible. These optical nanocircuits [19] can be designed to provide additional control of the space-charge fields and aberrations. Further research into generating enhanced beams using cathodes made of 'epsilon-near-zero' metamaterials or metamaterials with negative permittivity and permeability, called double-negative metamaterials [20] should be pursued.

This work opens a new direction for electron source research with direct applications in accelerator physics. Meta-cathodes like those described in this paper show the exciting prospect of controlling space charge forces at the cathode surface on the micron scale by custom engineering of the cathode's dielectric function.

**Acknowledgements**

I wish to thank SLAC for providing access to on-line journals which are behind pay walls. This work was privately funded.